\documentclass{aa}
\usepackage{graphicx}
\usepackage{txfonts}

\begin{document}

\title{The early spectral evolution of SN 2004dt}

\author{G.~Altavilla\inst{1,2} 
\and  M.~Stehle\inst{3} 
\and P.~Ruiz--Lapuente\inst{1,3}
\and P.~Mazzali\inst{3,4}
\and G.~Pignata\inst{5}
\and A.~Balastegui\inst{1} 
\and S.~Benetti\inst{6}
\and G.~Blanc\inst{6} \thanks{Now at APC, Universit\'{e} Paris 7, 10, rue Alice Domon et L\'{e}onie Duquet, 75205 Paris Cedex 13, France} 
\and R.~Canal\inst{1}
\and N.~Elias--Rosa\inst{6,7} \thanks{Now at Max--Planck--Institut f\"ur Astrophysik, Karl--Schwarzschild--Stra{\ss}e 1, D--85741 Garching,  Germany}
\and A.~Goobar\inst{8}
\and A.~Harutyunyan\inst{6}
\and A.~Pastorello\inst{3} \thanks{Now at the Astrophysics Research Centre, School of Mathematics and Physics, Queen's University Belfast, BT7 1NN, Belfast, United Kingdom} 
\and F.~Patat\inst{9} 
\and J.~Rich\inst{10} 
\and M.~Salvo\inst{11}
\and B.P.~Schmidt\inst{11}
\and V.~Stanishev\inst{8}
\and S.~Taubenberger\inst{3} 
\and M.~Turatto\inst{6}
\and W.~Hillebrandt\inst{3}
}
 
\offprints{G.~Altavilla}

\institute{
Department of Astronomy, University of Barcelona, Mart\'{\i}\ i 
Franqu\'es 1, E--08028 Barcelona, Spain. 
\and 
INAF--Osservatorio Astronomico di Bologna, Via Ranzani 1, I--40127 Bologna, Italy.
\\
\email{giuseppe.altavilla2@unibo.it}
\and 
Max--Planck--Institut f\"ur Astrophysik, Karl--Schwarzschild--Stra{\ss}e 1,
D--85741 Garching, Germany.
\and 
INAF--Osservatorio Astronomico di Trieste, via Tiepolo 11,
I--34131 Trieste, Italy
\and 
Departamento de Astronom\'{\i}a y Astrof\'{\i}sica, Pontificia Universidad 
Cat\'olica, Santiago de Chile, Chile
\and 
INAF--Osservatorio Astronomico di Padova, vicolo dell'Osservatorio 5, 
I--35122 Padova, Italy
\and 
Universidad de La Laguna, E--38206 La Laguna, Tenerife, Spain
\and 
Physics Department, Stockholm University, AlbaNova University Center,
106 91 Stockholm, Sweden
\and 
European Southern Observatory,  Karl--Schwarzschild--Stra{\ss}e 2, 
D--85748 Garching, Germany
\and 
Research School of Astronomy and Astrophysics, Australian National 
University, Cotter Road, Weston Creek, ACT 2611 Canberra, Australia
\and 
Australian National University, Mount Stromlo Observatory, Cotter Road,
Weston ACT 2611 Canberra, Australia
}

\abstract
{}
{We study the optical spectroscopic properties of Type Ia Supernova (SN Ia) 2004dt, focusing our attention on the early epochs.}
{Observation triggered soon after the SN 2004dt discovery allowed us to obtain a  spectrophotometric coverage  from day $-10$ 
to  almost one year ($\sim$353 days) after the $B$ band maximum. Observations carried out on an almost daily basis
allowed us a good sampling of the fast spectroscopic evolution of SN 2004dt in the early stages. To obtain this result,
low--resolution, long--slit spectroscopy was obtained using a number of facilities.}
{This supernova, which  in some absorption lines 
of its early spectra showed the highest degree of polarization ever measured in any 
SN Ia, has a complex velocity structure in the outer layers of its ejecta.
Unburnt oxygen is present, moving at velocities as high as $\sim$16,700 km s$^{-1}$,
with some intermediate--mass elements (Mg, Si, Ca) moving equally fast.
Modeling of the spectra based on standard density profiles of the ejecta
fails to reproduce the observed features, whereas enhancing the density of
outer layers significantly improves the fit. Our analysis indicates the presence
of clumps of high--velocity, intermediate--mass elements in the outermost
layers, which is  also suggested by the spectropolarimetric data.}
{}

\keywords{stars: supernovae: general - stars: supernovae: individual: SN 2004dt - method: observational - techniques: spectroscopic}

\maketitle

\section{Introduction}

Type Ia supernovae (SNe Ia) are thermonuclear explosions of white dwarfs (WDs) 
mostly originating from ignition of carbon close to the center 
of the star when it approaches the Chandrasekhar mass, due to accretion of 
matter from a close binary companion (see Hillebrandt \& Niemeyer \cite{hillebrandt2000} for a
review). The nature of the binary system and the mechanism of the explosion 
are still unsolved problems. Interest in solving these questions has strongly 
increased with the use of SNe Ia as distance indicators in cosmology, in 
particular by their role in the discovery of the acceleration of the 
expansion of the Universe (Riess et al. \cite{riess1998}; Perlmutter et al. \cite{perlmutter1999}). The 
nature of the companion (whether it is another WD 
---the double--degenerate or DD scenario--- or a non-degenerate star in any 
evolutionary stage ---the single--degenerate or SD scenario) is the  object 
of different approaches: searches for  DD systems with total mass 
exceeding the Chandrasekhar mass and able to merge in less than the  Hubble 
time (Napiwotzki et al. 2004); searches for semidetached SD systems where a 
WD close to the Chandrasekhar mass is growing by accretion (Maxted, Marsh, 
\& North \cite{maxted2000}); and finally searches for traces of hydrogen (whether deposited 
on the  surface of the exploding WD, 
residing in an accretion disk, stripped of the 
companion, or present as circumstellar material) in early  SNe Ia spectra
(Lundqvist et al. \cite{lundqvist2003}).  
A more direct approach, which is based on a possible  identification of 
the surviving companion of the WD in the central regions of the remnants of 
the Galactic historical SNe Ia,  would also
support the SD scenario (Ruiz--Lapuente et al. \cite{ruiz2004}).

Concerning the explosion mechanism, there is general agreement that explosive 
thermonuclear burning starts near the center of the WD and first propagates
subsonically outwards (deflagration). There is strong debate, though, on 
whether subsonic burning alone can produce both explosions with the typical 
SNe Ia energies and the chemical stratification revealed by the time sequence of 
SNe Ia spectra (avoiding, in particular, excessive mixing of unburnt carbon 
and oxygen within deep layers of the SN ejecta: see Reinecke, Hillebrandt \& 
Niemeyer \cite{reinecke2002}; 
Gamezo et al. \cite{gamezo2003}). Transition to detonation (supersonic burning propagated 
by a shock wave) has been invoked when burning reaches layers with densities low enough to 
burn only partially when they detonate (delayed detonation). This
takes place either in a spherical shell (Khokhlov \cite{khoklov1991}) or in a point (Livne 
\cite{livne1999}), but it has to be triggered artificially in current numerical 
simulations. Such detonations leave none or little of the
carbon and oxygen unburnt. 
Recent 2-D simulations have shown  a detonation  forming in the outer layers of an 
exploding WD when a hot plume rising from an off--center ignition spreads 
over the surface and sweeps the unburnt material that, when confined by the 
strong gravitational field, converges to a point (Plewa, Calder \& Lamb \cite{plewa2004}).
Nevertheless, subsequent 3-D simulations do not support this hypothesis, 
even if detonation, triggered by other mechanisms such as a pulsation or
spontaneous detonation if the turbolent energy is high enough, is not completely ruled out
(R\"{o}pke, Woosley \& Hillebrandt \cite{ropke2006}).

From the observational point of view, much  can be learned about the explosion mechanism 
from the detailed structure of  the layers that are just below the surface of the WD when it explodes. 
This  material  becomes rapidly   transparent as the SN ejecta expand, and a 
series of early--time spectra is thus required for studying  this phase properly.
A recent analysis of the distribution of the abundances in a large sample 
of SNe Ia  (Mazzali et al. \cite{mazzali2007}) suggests that the 
properties of most, if not all, SNe Ia 
can be explained by a single explosion scenario, the delayed detonation.
Pre--maximum spectropolarimetry of 17 SNe Ia analyzed by Wang, Baade \& Patat (\cite{wang2007})
strongly corroborates the delayed detonation model.

Spectra of SN 2004dt taken about 10 days before maximum light show a high 
degree of polarization across some lines (Wang et al. \cite{Wang06}), with Si\,{\sc ii} 
displaying the strongest polarization ($\sim2\%$) while O\,{\sc i} is hardly 
polarized. The  velocity structure is also complex (already noticed 
by Patat, Pignata \& Benetti \cite{patat2004}), with O\,{\sc i}, Mg\,{\sc ii}, Si\,{\sc ii}, 
and Ca\,{\sc ii} moving at high  velocities\footnote{A distinction should be made
between  high--velocity features that persist through the  optical maximum and the
``high--velocity features" (HVFs), as the   the Ca\,{\sc ii} IR triplet, commonly seen 
at very  early epochs ($\sim 10$ days before maximum  light).
The high velocities characterizing these features,
much higher than the photospheric velocities, suggest that these lines (or a component 
of the photospheric line if not detached)  
originate above the photosphere and that they are a different 
phenomena from to the normal photospheric absorption features.
 As discussed by Tanaka et al. 
(\cite{tanaka06}),  HVFs  may be the result of density and/or abundance  
enhancements produced by the explosion itself or by the interaction of the SN with the 
circumstellar material (in this hypothesis high mass--loss rates are required for the progenitor) or 
with an  accretion disk (the  absence of hydrogen lines in the early spectra weaken this
model, but see Gerardy et al. \cite{gerardy2004} and Mazzali et al. \cite{Mazza05b}). 
The Ca\,{\sc ii} IR HVF is nicely explained within the gravitationally confined detonation (GCD) 
model by Kasen \& Plewa (\cite{kasen2005}).
}
while S\,{\sc ii}
and Si\,{\sc iii} appear significantly slower.  This,  together with the polarization 
data, suggests the presence of an almost spherical O layer with intrusions of 
intermediate--mass elements. While the finger--like structure is typically 
found in pure deflagration models of the explosions, Kasen \& Plewa (\cite{kasen2005})
interpret the presence of Ca\,{\sc ii} moving at high velocities as a signature of 
the deflagration products brought to the surface by the rising plume in the 
model of Plewa, Calder \& Lamb (\cite{plewa2004}), which are compressed and accelerated 
by the SN ejecta after the detonation has been triggered.   
\\
The expansion velocity  of the strong Si\,{\sc ii} absorption, 
starting from $\sim$16,000 km s$^{-1}$ at about
10 days  maximum light and  decreasing fast with time, also  suggests that
SN 2004dt is a high--velocity gradient (HVG) event (see Benetti et al. \cite{Bene05}),
as confirmed by a preliminary photometric analysis, and similar to other
SNe  such as SN 2002bo (Benetti et al. \cite{benetti2004}), 
its twin SN 2002dj (Pignata et al. \cite{pignata2005}), SN 2002er (Pignata et al. \cite{pignata2004};
Kotak et al. \cite{kotak2005}), and  SN 2005cf (Pastorello et al. \cite{pastorello2007}; 
Garavini et al. \cite{garavini2007}), which
have been extensively observed within the same
scientific project: ``The Physics of Type Ia Supernova Explosions, a European
Research Training Network''\footnote{http://www.mpa-garching.mpg.de/$\sim$rtn/}
(RTN), 
carried out by the  European Supernova Collaboration (ESC,  Benetti 
et al. \cite{benetti2004}).

Here we present a series of  optical spectra of SN 2004dt, 
from more than one week before maximum light to almost one year after, focusing our attention
on spectra obtained at early epochs. This is a subsample 
of the whole data set collected for SN 2004dt
on behalf of the  RTN program. The complete data set includes
optical and infared  photometric data (UBVRI photometry, $58$ nights, from -10 to +500
days since $B$ maximum;
JHK photometry, $18$ nights, from -8 to 357 days since $B$ maximum); 
and optical and  infrared spectroscopy ($32$ nights in the optical range, from -10 to
353 days  and $5$ nights in IR, from -5 to +16 days since $B$ maximum), 
which will be analyzed and discussed in a forthcoming paper.
\\
We examine here the peculiarities
of SN 2004dt as inferred from the shape and velocities of the spectral
features before maximum and attempt to model the spectra from 1-D 
hydrodynamical models of the explosion. We find that the best fits are
obtained when we artificially raise the density of the outermost layers,
reinforcing the idea that both the velocity structure and the line
polarization observed are due to clumps of intermediate--mass elements
on and above the ``photosphere'' of the SN.

\begin{figure}[t]
\includegraphics[angle=0,width=\columnwidth]{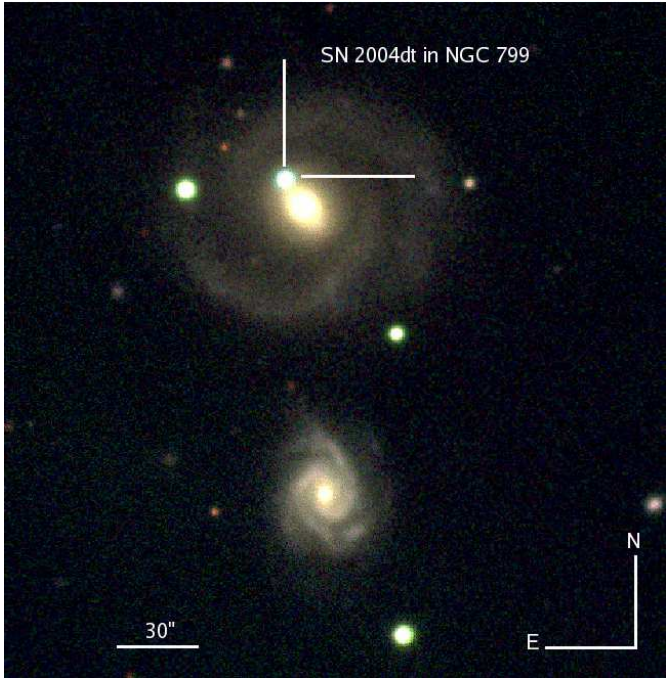}
\caption{SN 2004dt in NGC799. Color image obtained by combining B,V,R images
(80, 50, 40 sec exposure each) obtained on Aug. 19th with the 2.2m+Cafos at 
Calar  Alto Observatory. The dashes indicate the position of the supernova.}
\label{SN2004dtGalaxycolour}
\end{figure}
\begin{table}[t]
\caption{Main parameters or SN 2004dt and its host galaxy}
\label{table:1}
 \setlength\tabcolsep{5pt}
\begin{tabular}{@{}ll}
\\
\hline\hline
Parent galaxy: & NGC 799\\
Galaxy type:   & (R')SB(s)a $^{\star}$\\
Heliocentric Radial Velocity: &  5915 km s$^{-1}$$^{\star}$ \\
Radial velocity (vvir): &  5846 km s$^{-1}$$^{\ddag}$ \\
Redshift:               &  0.01973 $^{\dag}$ \\
Kinematical distance modulus: & 34.61$^{\bullet}$\\
R.A.$_{SN}$ (J2000): & 02$^{h}$02$^{m}$12$^{s}$.77\\
Dec.$_{SN}$  (J2000): & -00$^{\circ}$ 05' 51''.5\\
Offset from nucleus: &7".0 E, 10".5 N\\
Discovery epoch: & JD=2453228.97(Aug. 11$^{th}$)\\
Date of B maximum: & JD=$2453240.3 \pm 0.5$ (Aug. 22$^{nd}$)\\
Observed magnitude at max: &  B$= 15.33 (0.02)$\\
$\Delta m_{15}(B)_{obs}$:& $ 1.21 (0.05)$\\
Stretch factor (B): & $0.91 (0.01)$\\
Galactic Extinction: & A$_{B}$ = 0.010 mag$^{\diamond}$;  A$_B$ = 0.109 mag$^{\triangleright}$\\ 
(B-V)$_{max}$:& $-0.03\pm0.02^{\triangleleft}$\\
E(B-V)$_{max}$:& $0.04\pm0.04^{\triangleleft}$\\
E(B-V)$_{tail}$:&$0.11\pm0.03^{\triangleleft}$\\
\hline
\hline
\end{tabular}
\begin{list}{}{}
\item[$^{\star}$] NED (Nasa/ipac Extragalactic Database)
\item[$^{\ddag}$] Leda (Lyon-Meudon Extragalactic Database), corrected for LG infall onto Virgo
\item[$^{\dag}$]  Theureau et al. \cite{theureau1998}
\item[$^{\bullet}$] Leda (Lyon-Meudon Extragalactic Database), from vvir and Ho=70 km s$^{-1}$ Mpc$^{-1}$
\item[$^{\diamond}$] Burstein \& Heiles \cite{burstein1982}
\item[$^{\triangleright}$] Schlegel et al. \cite{schlegel} (we make use of this value)
\item[$^{\triangleleft}$] Galactic extinction correction applied
\end{list}
\end{table}

\section{Observations and data reduction}

SN 2004dt (RA: 02$^h$02$^m$12$^s$.77, Dec: -00$^{\circ}$05'51".5, J2000)  
was discovered on August 11.48 UT by  Moore \& Li (\cite{moore2004})
on behalf of the LOSS/KAIT\footnote{Lick Observatory Supernova Search
with the  Katzman Automatic Imaging Telescope, \\
http://astro.berkeley.edu/$\sim$bait/kait.html  supernova search} (Filippenko et al. \cite{filippenko2001}),
in the spiral galaxy NGC 799, 7".0 East 
and 10".5 North  from the galaxy nucleus
(see Fig. \ref{SN2004dtGalaxycolour} and Table \ref{table:1} for the main
parameters of SN 2004dt and its host  galaxy).
\\
Spectra of SN 2004dt were obtained immediately on 
August 12.72 UT at  the ANU 2.3-m telescope (Salvo, Schmidt \& Wood \cite{salvo2004}) 
and on August  13.17 UT at the Calar Alto  2.2-m telescope 
(Patat et al. \cite{patat2004}).  
On the basis of these spectra, the candidate was classified 
as a young Type Ia with some unusual features 
(Salvo et al. \cite{salvo2004}; Patat et al. \cite{patat2004}). 
The proximity of the host galaxy (z=0.01973,  Theureau et al. \cite{theureau1998}),  
the early epoch of the discovery, the  peculiar spectral
 features, the  good position of  the SN,  easily observable from both the 
northern and  southern observatories for a long period (the host galaxy had 
just  appeared from behind the Sun),
made SN 2004dt a good target for the ESC. For these reasons 
target-of-opportunity  (ToO)  observations were immediately triggered.  
Later on, observations were  carried out either in 
service or in target-of-opportunity mode, using the configuration  available at the requested time.
Spectroscopy and imaging were carried out at several 
sites using a number of different instrumental configurations. 
Table \ref{table2} contains the log of the spectroscopic observations.

Data reduction was performed using standard 
{\sc IRAF}\footnote{http://iraf.noao.edu/ \\ 
{\sc IRAF} is distributed by the National Optical Astronomy Observatories 
(NOAO),  which are operated by the Association of Universities for Research 
in Astronomy (AURA),  Inc., under cooperative agreement with the National 
Science Foundation.}
 procedures for long--slit spectroscopy. 
All images were bias--subtracted and then flat-field corrected using dome flats. 
The background was interpolated by fitting the region on both sides of the spectrum 
with a low--order polynomial and then subtracted. 
Extractions were usually weighted by the variance based on the data values and 
a Poisson/CCD  model using the gain and readout noise parameters\footnote{
Variance  weighting  is  often  called  ``optimal"  extraction  since  it 
produces  the  best unbiased signal-to-noise estimate of the flux in the
2-D profile (Horne \cite{horne1986}; Marsh \cite{marsh1989}).}.

Spectra were wavelength--calibrated by means of reference arc spectra,  and the 
resulting wavelength--calibrated spectra were checked by measuring the position  of known bright--night  skylines
(usually the [OI] lines at $\lambda 5577$ \AA\ and $\lambda 6300$ \AA). 
Small discrepancies have been corrected by applying a rigid shift in order
to match the skylines wavelengths.
The spectra were flux--calibrated using a spectroscopic standard star observed 
with the same telescope setup and reduced as was the supernova itself. 
If spectrophotometric  standard stars were not  available for a given epoch, 
response curves  obtained for the same  telescope/instrument
in another night close in time were used.
The flux calibration was then   checked against the
photometry and, when necessary, spectral fluxes were scaled to match 
the photometric data.
Partial correction for atmospheric absorption was also applied. 
Multiple spectra of the SN obtained in the same night  were combined 
in order to improve  the signal-to-noise  ratio and  almost simultaneous spectra 
covering different but overlapping spectral ranges were merged  together  in a single, 
more extended spectrum.
When the observing conditions were not photometric, spectra covering different wavelength 
ranges were scaled  to match in the  overlapping region.

\begin{table*}[!ht]
\caption{Log of the optical spectroscopic observations of SN 2004dt}
\label{table2}
  \centering
\begin{tabular}{@{}llllllr}
\\
\hline\hline
Date & \multicolumn{1}{c}{JD} & Epoch$^{\star}$ & Approx. range   & Telescope& Flux & Reference\\
\multicolumn{1}{c}{}    & \multicolumn{1}{c}{-2400000} & \multicolumn{1}{c}{[days]}  &  
\multicolumn{1}{c}{[\AA]}      &    \multicolumn{1}{l}{+Instrument$^*$}    & \multicolumn{1}{l}{standard} & \\
\hline
2004/08/12 & 53230.22 & -10.1 & 4500-10300 & ANU2.3+DBS    &  hr718     & M. Salvo\\
2004/08/12 & 53230.67 &  -9.6 & 3200-8600  & CA2.2+CAFOS   &  bd+284211 & G. Pignata\\
2004/08/13 & 53231.23 &  -9.1 & 4000-9700  & ANU2.3+DBS    &  hr718     & M. Salvo\\
2004/08/13 & 53231.63 &  -8.7 & 3250-9150  & CA2.2+CAFOS   &  bd+284211 & G. Pignata\\
2004/08/15 & 53233.64 &  -6.7 & 3200-10000 & CA2.2+CAFOS   &  bd+284211 & G. Pignata\\
2004/08/15 & 53233.69 &  -6.6 & 3600-8900  & NOT+ALFOSC    &  Feige110  & V. Stanishev\\
2004/08/16 & 53233.71 &  -6.6 & 3100-10200 & TNG+DOLORES   &  -         & N. de la Rosa\\
2004/08/16 & 53234.57 &  -5.7 & 3200-9600  & CA2.2+CAFOS   &  bd+284211 & G. Pignata\\
2004/08/17 & 53235.69 &  -4.6 & 3300-9700  & NOT+ALFOSC    &  bd+284211 & V. Stanishev\\
2004/08/18 & 53236.54 &  -3.8 & 3200-9800 & CA2.2+CAFOS   &  bd+284211 & G. Pignata\\
2004/08/19 & 53237.18 &  -3.1 & 3300-9000  & ANU2.3+DBS    &  Feige110  & M. Salvo\\
2004/08/20 & 53238.22 &  -2.1 & 3300-8800  & ANU2.3+DBS    &  Feige110  & M. Salvo\\
2004/08/21 & 53239.23 &  -1.1 & 3300-8800  & ANU2.3+DBS    &  Feige110  & M. Salvo\\
2004/08/21 & 53239.59 &  -0.7 & 3400-7700  & ASIAGO+AFOSC  &  bd+254655 & S. Benetti\\
2004/08/23 & 53241.28 &  +1.0 &  3500-7200  & ANU2.3+DBS    &  Feige110  & M. Salvo\\
2004/08/24 & 53242.28 & 2.0 & 3500-7200 & ANU2.3+DBS    &  Feige 110     & M. Salvo\\
2004/08/24 & 53242.59 & 2.3 & 3300-9000 & NOT+ALFOSC    &  bd+284211     & V. Stanishev\\
2004/08/25 & 53243.68 & 3.4 & 3300-9100 & CA2.2+CAFOS   &  bd+284211     & G. Pignata\\
2004/08/26 & 53244.15 & 3.8 & 3250-9150 & CA2.2+CAFOS   &  bd+284211     & G. Pignata\\
2004/08/31 & 53249.58 & 9.3 & 3300-7900 & TNG+DOLORES   &  HR9087        & N. de la Rosa\\
2004/09/05 & 53254.58 &14.3 & 3350-9000 & NOT+ALFOSC    &  -             & V. Stanishev\\ 
2004/09/08 & 53257.63 &17.3 & 3400-7700 & ASIAGO+AFOSC  &  bd+254655     & N. de la Rosa\\
2004/09/12 & 53260.69 &20.4 & 3500-8900 & NOT+ALFOSC    &  -             & V. Stanishev\\ 
2004/09/18 & 53267.19 &26.9 & 3500-9000 & ANU2.3+DBS    &  Feige 110     & J. Rich \\ 
2004/09/19 & 53268.15 &27.9 & 3500-9000 & ANU2.3+DBS    &  Feige 110     & J. Rich \\ 
2004/09/28 & 53277.58 &37.3 & 3400-9000 & NOT+ALFOSC    &  -             & V. Stanishev\\ 
2004/10/08 & 53287.48 &47.2 & 3500-9200 & CA2.2+CAFOS   &  Feige 110     & G. Pignata\\
2004/10/22 & 53301.09 &60.8 & 3400-9000 & ANU2.3+DBS    &  Feige 110     & M. Salvo \\ 
2005/01/21-22& 53392.80&152.5&3500-8600 & CA2.2+CAFOS   &  Feige 34      &A. Pastorello\\
             &         &     &          &               &  G191-B2B      &S. Taubenberger\\
2005/08/11 & 53593.78 &353.5 &4000-8400 & VLT+FORS1     &  GD50          &N. de la Rosa\\
             &         &     &          &               &                &S. Benetti\\
\hline\hline
\end{tabular}\\[2pt]
\begin{list}{}{}
\item[$^{\star}$] relative to the  B band maximum.
\item[$^*$] ANU: Australian National University 2.3-m telescope,
DBS: Double Beam Spectrograph;  
CA2.2: Calar Alto 2.2-m telescope,
CAFOS: Calar Alto Faint Object Spectrograph;
NOT: Nordic Optical Telescope (2.5-m),
ALFOSC Andalucia Faint Object Spectrograph and Camera;
TNG: Telescopio Nazionale Galileo (3.6-m),
DOLORES:  Device Optimized for the LOw RESolution,
ASIAGO: 1.8-m Copernico telescope
AFOSC: Asiago Faint Object Spectrograph and Camera,
VLT: ESO Very Large Telecope, Unit Telescope 2 (8.2m),
FORS1: FOcal Reducer/low dispersion Spectrograph. 
\end{list}
\end{table*}

\begin{figure}[hbtp]
\includegraphics[angle=0,width=\columnwidth]{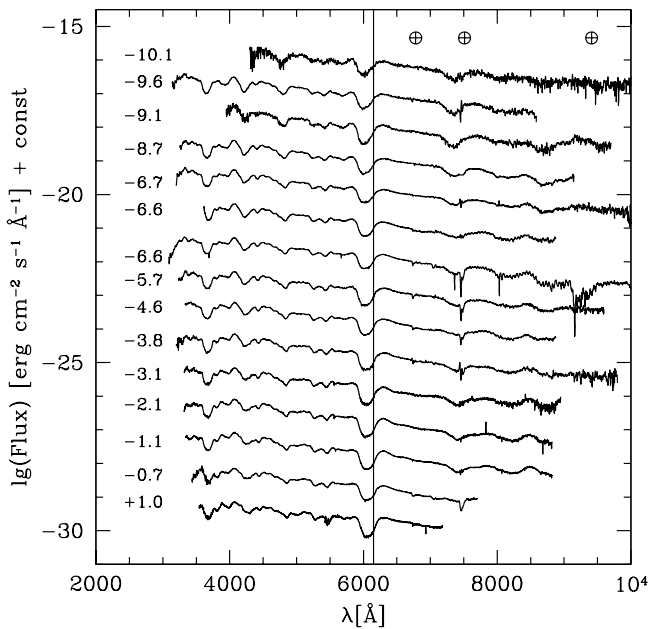}
\caption{Early spectral sequence for SN 2004dt. 
The vertical line at $\lambda=6150$\,\AA\,  makes
the evolution of the blueshifted Si\,{\sc ii} $\lambda 6355$\,\AA\, feature 
clear. 
Spectra are restframe. 
No reddening correction has been applied.
The $\oplus$ symbols mark the position of the main telluric absorptions, 
which have been  partially removed. Spectra have been shifted downwards  
for clarity (shift step = -1). 
The phase with respect to the  maximum of the B light curve 
is shown to the left of each spectrum.}
\label{SN2004dtspec}
\end{figure}

\begin{figure}[hbtp]
\includegraphics[angle=0,width=\columnwidth]{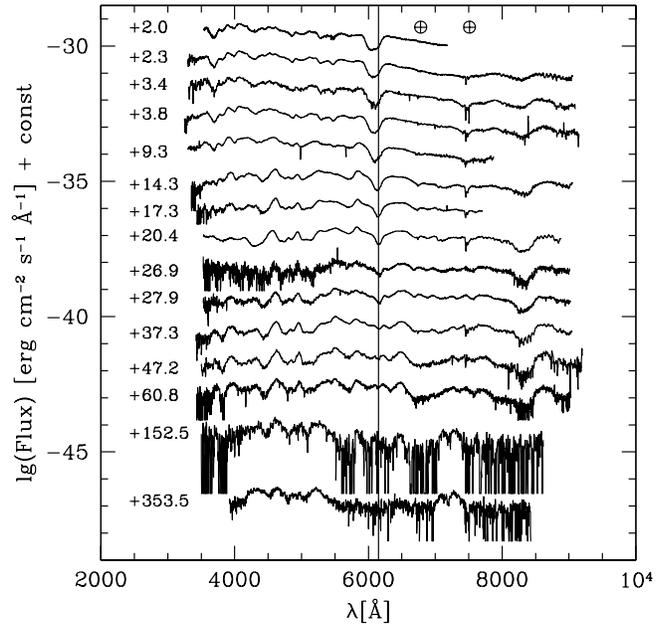}
\caption{SN 2004dt post maximum  spectral evolution. 
Labels as in Fig. \ref{SN2004dtspec}.}
\label{SN2004dtspec2}
\end{figure}

There was  good pre--maximum spectrophotometric
coverage, which is  one of the main goals of the European Supernova Collaboration. 
In fact, soon after the explosion, Type Ia
SNe show  differences that are bigger than those at later phases, hence early observations
can provide useful hints for  understanding the nature of the progenitors and the explosion mechanism. 
The spectral evolution of SN 2004dt from day -10 to day +1 is shown in
Fig. \ref{SN2004dtspec}, while the postmaximum spectral  evolution
is shown in Fig. \ref{SN2004dtspec2}. 
The good temporal coverage allows 
us to follow the  fast evolution at early times on a daily basis.
Early spectra  exhibit an unusually broad and asymmetric 
Si\,{\sc ii} $\lambda 6355$\,\AA\ feature, analogous with SN 2005cf (Garavini et al. \cite{garavini2007}), 
which gradually shifts to the red and  
becomes  narrower at later phases. Strong absorption features identified
as Ca\,{\sc ii}, S\,{\sc ii}, Mg\,{\sc ii}, and O\,{\sc i}  are also evident. 
A more careful analysis shows that   almost all
absorption troughs, except for the S\,{\sc ii}  
$\lambda\lambda$5454, 5640\AA,  and Si\,{\sc iii} $\lambda 4565$\,\AA,
are highly blueshifted, as discussed 
in details in  Sect. \ref{tomography}.
A possible line identification 
coming from the
spectral modeling is  shown in Sect. \ref{standarddensityprofile}. 

\section{Light--curve parameters}
\label{colourevolution}
The photometric evolution of SN 2004dt will be discussed in detail in a forthcoming paper, while here
we report  some preliminary estimates of the relevant  photometric parameters.
A preliminary analysis of the 
optical  light curves
gives $B_{max}=15.33\pm0.02$
on Aug. $22^{nd}$ ($JD=2453240.3 \pm 0.5$). 
The decline rate after maximum is 
$\Delta m_{15}(B)= 1.21\pm 0.05$.
After the correction for the Galactic extinction ($A_B=0.109$ mag, as 
measured by  Schlegel, Finkbeiner \& Davis \cite{schlegel}), we obtain
$(B-V)_{max}=   -0.03\pm0.02$, corresponding to $E(B-V)_{max}= 0.04\pm 0.04$, 
which is  consistent with negligible reddening. This is derived from
the relation 
$(B-V)_{max}=0.09(\pm 0.08)\times(\Delta m_{15}-1.1)-0.08(\pm 0.03)$, 
(Altavilla et al. \cite{altavilla04}), which gives $(B-V)_{max}=-0.07\pm0.04$.
According to the Lira relation (Lira \cite{Lira1995}), we find
$E(B-V)_{tail}= 0.11\pm0.03$.  
The  weighted average color
excess is  $E(B-V)_{avg}= 0.08\pm0.05$.
The discrepancies between the two values of the color excess 
can be due to the SN spectroscopic  peculiarities that affect the 
color evolution and make the usual reddening law fail or lose accuracy. 
The absence of   clearly detectable  Na\,{\sc i}\ D lines in the spectra supports the hypothesis of a 
small reddening, if any. In this case
the absolute B magnitude at maximum would
be $B_{max}\sim-19.4$ ($H_0=70$ km s${^1}$ Mpc$^{-1}$).

\section{Expansion velocities and spectral properties  derived from early 
spectra}\label{tomography}
\begin{figure}[hbtp]
\includegraphics[angle=0,width=\columnwidth]{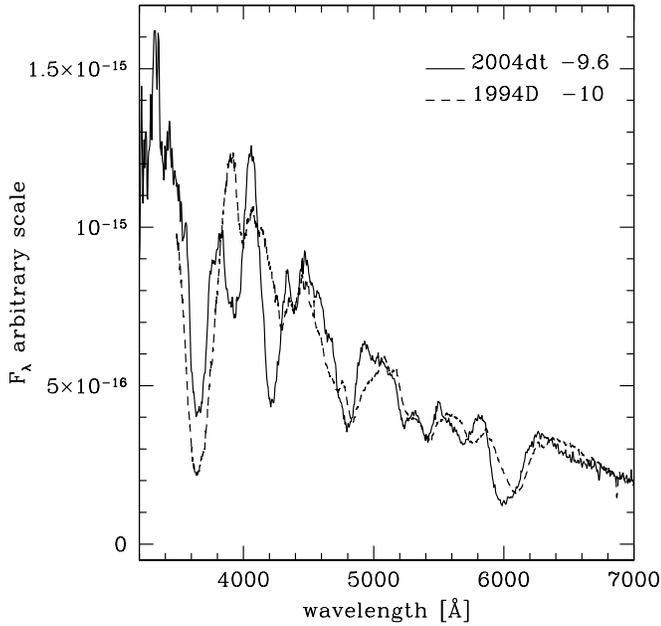}
\caption{Comparison among spectra of  SN 2004dt (solid line) and SN 1994D 
(dashed line) $\sim10$ days before maximum. To be noticed the very
different blue--shift of the Si\,{\sc ii} $\lambda$6355\AA\ features and the more
similar blue--shift of  the S\,{\sc ii} $\lambda$5640\AA\ lines. Significant differences are
present in the 3800-4200\AA\ region.
}
\label{SN2004dt1994D}
\end{figure}

\begin{figure}[hbtp]
\includegraphics[angle=0,width=\columnwidth]{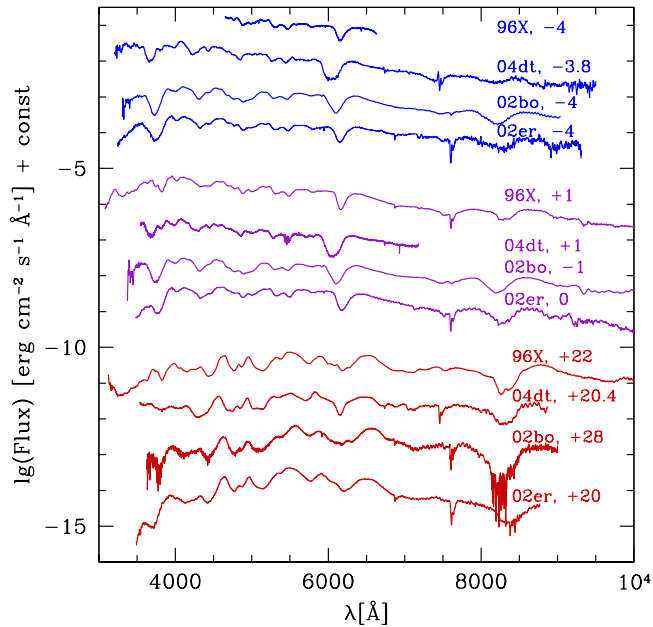}
\caption{Comparison among spectra of SN 2004dt, SN 1996X, SN 2002bo and SN 2002er 
at three different epochs (premaximum, maximum, postmaximum).}
\label{spec04dt96X}
\end{figure}

By looking at the early spectra, the peculiar large,  
asymmetric and  flat--bottomed  Si\,{\sc ii} $\lambda 6355$\,\AA\ feature is immediately evident.
The comparison of the SN 2004dt spectrum 10 days before maximum with SN 1994D
(Patat et al. \cite{patat1996})
at the same epoch shows the peculiarity of the SN 2004dt spectrum and
a high blue shift of the  Si\,{\sc ii} $\lambda 6355$\,\AA\ becomes evident, too
(Fig. \ref{SN2004dt1994D}). 
The peculiarities of SN 2004dt also persist  later on, 
as shown in Fig. \ref{spec04dt96X}, where spectra of SN 2004dt at different epochs 
are compared with those  of the normal SN Ia 1996X (Salvo et al. \cite{salvo2001}) and with
other two  HVG SNe: SN 2002bo (Benetti et al. \cite{benetti2004})
 and SN 2002er (Pignata et al. \cite{pignata2004}; Kotak et al. \cite{kotak2005}).

The analysis of the very early (epoch $\sim -10$)
spectrum gives the following results.
A strong  absorption,  measured at $\lambda 3930$\,\AA,  was initially
identified  as C\,{\sc ii} $\lambda 4267$\,\AA\ (Mazzali \cite{mazzali2001}; Branch et al. \cite{branch2003}; Thomas et al. \cite{thomas2007}),  
implying an expansion 
velocity of   about 23,700 km s$^{-1}$. In this hypothesis the other 
intense C\,{\sc ii} $\lambda\lambda 6576-6583$\,\AA\ transitions are blended 
with the Si\,{\sc ii} $\lambda 6355$\,\AA, which indeed appears very broad and intense
and has an asymmetric absorption trough.  On the other hand, there was no 
strong evidence of the  C\,{\sc ii} $\lambda\lambda 7231,7236$\,\AA\ line 
(this line, however, is expected to be much weaker, Mazzali \cite{mazzali2001}).
A second hypothesis was that the $\lambda 3930$\,\AA\ feature was a strong
absorption typically attributed to
Si\,{\sc ii} $\lambda\lambda$4128,4131\AA\ (Pskovskii \cite{pskovskii1969}), 
implying an expansion  velocity of  about 14,500 km s$^{-1}$.
Synthetic spectra (see Sect. \ref{density}) can reproduce the observations without using C, thus  rejecting
the first hypothesis and supporting the Si line identification.

\begin{figure}[hbtp]
\includegraphics[angle=0,width=\columnwidth]{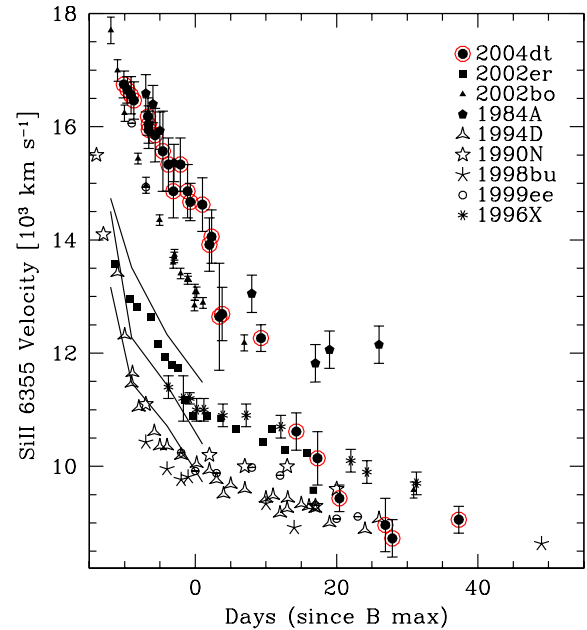}
\caption{The Si\,{\sc ii}  $\lambda 6355$\,\AA\ expansion  velocity evolution
 as deduced from the position of the  absorption  minimum, compared with other SNe and with
the expected evolution for different metallicities (solid line, top:
$\times 10$ solar metallicity; middle: $\times 1$;
 bottom: $\times 0.1$) obtained by
Lentz et al. (\cite{lentz2000}).
}
\label{SiIIVelocity}
\end{figure}

\begin{figure}[hbtp]
\includegraphics[angle=0,width=\columnwidth]{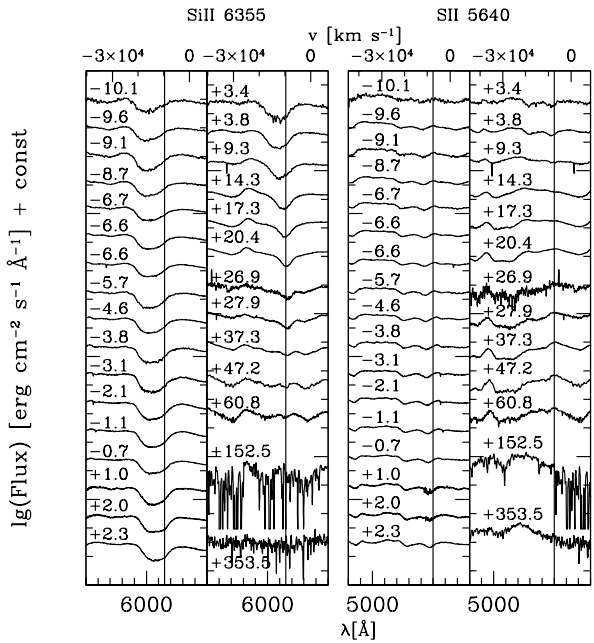}
\caption{Evolution with time of the Si\,{\sc ii}  $\lambda 6355$\,\AA\ (left panel) and 
S\,{\sc ii} $\lambda\lambda 5454,5640$\,\AA\ (right panel).
Vertical lines at 6150 and 5500\AA\ are shown to guide the eye.
}
\label{SiIIspecplotcal}
\end{figure}

The Si\,{\sc ii} $\lambda 6355$\,\AA\ minimum at ~6000\AA\ gives an expansion velocity 
of about 16,700 km s$^{-1}$ and  the O\,{\sc i} $\lambda 7773$\,\AA\ line 
(whose minimum is at ~7340\AA) also gives a similar expansion  velocity of 
16,700 km s$^{-1}$. The Si\,{\sc ii} $\lambda 6355$\,\AA\ expansion velocity
measured for SN 1994D at similar epochs (Patat et al. \cite{patat1996}) is 
significantly lower: 
$\sim 14,000$ km s$^{-1}$ at -11 days and  
$\sim 13,000$ km s$^{-1}$ at -10 days.
The  Ca\,{\sc ii} H\&K line is characterized by a velocity of about 22,500 km s$^{-1}$,
equal to the   Ca\,{\sc ii} H\&K velocity measured for SN 1994D at day -10.
In Fig. \ref{SiIIVelocity}, the Si\,{\sc ii} 
$\lambda 6355$\,\AA\ velocity evolution is shown and compared
with those of other SNe~Ia. The fast decline in the
Si\,{\sc ii} velocity is evident, and corresponds to a decline rate of
$\dot v=160\pm15$ km s$^{-1}$ d$^{-1}$, where $\dot v=-\Delta v/\Delta t$ is the average daily 
rate of decrease in the expansion velocity after maximum (Benetti et al. \cite{Bene05}). 
Lentz et al. (\cite{lentz2000}) computed the Si\,{\sc ii}  $\lambda 6355$\,\AA\  expected velocity 
evolution in SNe~Ia by varying the metallicity in the C+O layer in the standard 
deflagration
 model  W7 (Nomoto, Thielemann \& Yokoi \cite{nomoto1984}; Thielemann, Nomoto, \& Yokoi 
\cite{thielemann1986}). The expected trends for 3 different metallicities are also shown in  
Fig. \ref{SiIIVelocity}. The high expansion velocity of  SN 2004dt, as well
as those of SN 2002bo or SN 1984A, cannot be reproduced even with  unreasonable
high metallicity values, since the unmixed W7 model predicts no Si at velocities 
higher than $\sim 15,000 $ km s$^{-1}$.
The evolution in time of the Si\,{\sc ii} 
$\lambda 6355$\,\AA\ line shape is shown in Fig. \ref{SiIIspecplotcal}.
The line shows a wide flat--bottomed profile until some days past maximum,
the first spectrum with a narrower profile being the one taken at about 2
weeks past maximum.

\begin{figure}[hbtp]
\includegraphics[angle=0,width=\columnwidth]{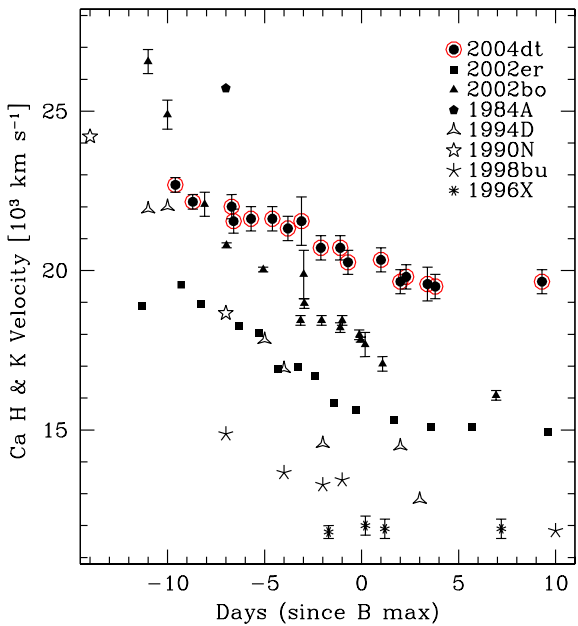}
\caption{The Ca\,{\sc ii} H \& K  expansion  velocity as deduced from 
the position of its minimum, compared with those of other SNe.}
\label{CaHKVelocity}
\end{figure}
In Fig.  \ref{CaHKVelocity} the  Ca\,{\sc ii} H\&K velocity evolution is shown
and compared with those of other SNe~Ia.
However, the Ca\,{\sc ii} H\&K feature could be  dominated by
the blended Si\,{\sc ii} $\lambda\lambda$3856,3863\AA\ line, while the 
Si\,{\sc ii} $\lambda 4130$\,\AA\ is well--separated.  
In this case, the corresponding Si\,{\sc ii}  $\lambda\lambda$3856,3863\AA\
velocity is significantly lower ($\leq 16,150$ km s$^{-1}$) than
the one computed for Ca\,{\sc ii} H\&K,  and more like
to the Si\,{\sc ii} $\lambda 6355$\,\AA\ values, thus supporting this hypothesis.
Moreover,  the comparison of the spectra with a theoretical model (discussed in Sect. \ref{density})
reinforces  this second  hypothesis.
Ca\,{\sc ii} H\&K contamination by Si\,{\sc ii} has already been suggested by
Kirshner et al. (\cite{kirshner1993}), who pointed out that the Si\,{\sc ii} $\lambda 3858$
is fainter, but comparable, to the Ca\,{\sc ii}  H\&K feature, and
 by Nugent et al. (\cite{nugent1997}), 
who suggested  that the ``split'' just blueward of the Ca\,{\sc ii} H \& K 
feature visible in both the observed  spectrum at  maximum  of SN 1994D  and 
the corresponding W7 synthetic spectra,
could  likely be  explained by  a blend with  Si\,{\sc ii} $\lambda 3858$.
Lentz et al. (\cite{lentz2000}) confirmed the identification of the blue wing with the 
Si\,{\sc ii}  line. Several other SNe, as for example SN 1990N, SN 1996X, and 
probably SN 1998bu, and SN  2003du show this 
contamination, too (Stanishev et al. \cite{stanishev2006}).

\begin{figure}[hbtp]
\includegraphics[angle=0,width=\columnwidth]{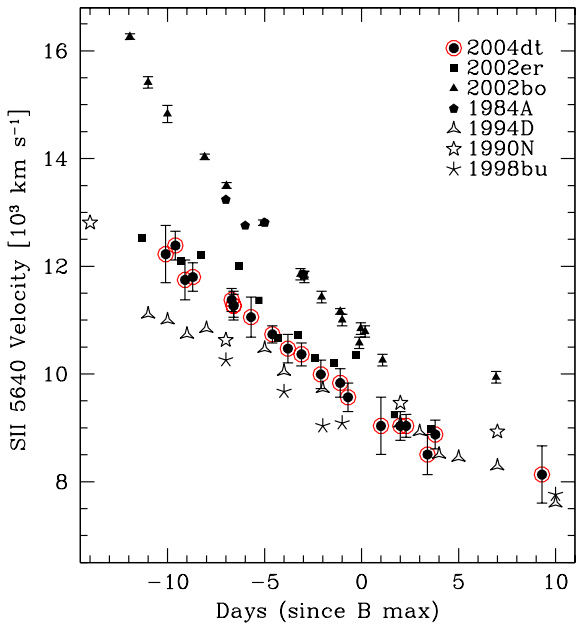}
\caption{The S\,{\sc ii} $\lambda 5640$\,\AA\  expansion  velocity as deduced from 
its minimum, compared with those of other SNe.}
\label{SIIVelocity}
\end{figure}

\begin{table}[t]
\caption{Expansion velocities from  the Si\,{\sc ii} 6355\AA, 
Ca\,{\sc ii} H\&K, or Si\,{\sc ii} 3860\AA\  and S\,{\sc ii} 5640\AA\ absorption}
\label{velocity}
  \centering
\begin{tabular}{@{}l r r r r  }
\\
\hline\hline
Phase & Si\,{\sc ii} 6355 & H\&K$^{\bullet}$  &  Si\,{\sc ii} 3860$^{\bullet}$ & S\,{\sc ii} 5640\\
 $[$day$]$& [km s$^{-1}$] & [km s$^{-1}$] & [km s$^{-1}$] & [km s$^{-1}$]  \\

\hline
   -10.1     &  16750 (235)  &      -       & 		   &   12230 (531)\\
    -9.6     &  16650 (235)  & 22690 (227)  & 16150 (233)  &   12390 (266)\\
    -9.1     &  16560 (330)  &	    -       & 15610 (233)  &   11750 (372)\\
    -8.7     &  16460 (330)  & 22160 (227)  &      -	   &   11800 (266)\\
    -6.7     &  16180 (330)  & 22000 (379)  & 15460 (388)  &   11380 (213)\\
    -6.6     &  15940 (330)  & 21550 (379)  & 14990 (388)  &   11270 (213)\\
    -6.6     &  16040 (330)  &	    -       &      -	   &   11270 (266)\\
    -5.7     &  15850 (471)  & 21630 (379)  & 15070 (388)  &   11060 (372)\\
    -4.6     &  15570 (707)  & 21630 (379)  & 15070 (388)  &   10740 (160)\\
    -3.8     &  15330 (471)  & 21320 (379)  & 14760 (388)  &   10470 (266)\\
    -3.1     &  14860 (471)  & 21550 (758)  & 14990 (777)  &   10370 (213)\\
    -2.1     &  15330 (471)  & 20710 (379)  & 14140 (388)  &   9993  (266)\\
    -1.1     &  14860 (471)  & 20710 (379)  & 14140 (388)  &   9834  (266)\\
    -0.7     &  14670 (330)  & 20260 (379)  & 13670 (388)  &   9568  (266)\\
    +1.0     &  14620 (471)  & 20340 (379)  & 13750 (388)  &   9036  (532)\\
    +2.0     &  13920 (471)  & 19650 (379)  & 13050 (388)  &   9036  (266)\\
    +2.3     &  14060 (471)  & 19800 (379)  & 13200 (388)  &   9036  (213)\\
    +3.4     &  12640 (943)  & 19580 (531)  & 12970 (543)  &   8505  (372)\\
    +3.8     &  12690 (471)  & 19500 (379)  & 12890 (388)  &   8877  (266)\\
    +9.3     &  12270 (235)  & 19650 (379)  & 13050 (388)  &   8133  (532)\\
    +14.3    &  10610 (330)  &	-	    &	-	   &	- \\
    +17.3    &  10140 (471)  &	 -	    &	-	   &	- \\
    +20.4    &  9435  (235)  &	  -         &	-	   &	- \\
    +26.9    &  8963  (471)  &	   -	    &	-	   &	- \\
    +27.9    &  8727  (330)  &	    -	    &	-	   &	- \\
    +37.3    &  9057  (235)  &	     - 	    & 	-	   &	- \\
\hline\hline
\end{tabular}\\
\begin{list}{}{}
\item[$^{\bullet}$] see discussion is Sect. \ref{tomography}
\end{list}
\end{table}
SN 2004dt expansion velocities for Si\,{\sc ii} $\lambda 6355$\,\AA, Si\,{\sc ii} 
$\lambda\lambda$3856,3863\AA,
and Ca\,{\sc ii} H\&K  are tabulated in Table \ref{velocity}. The large uncertainties
are mainly due to the wide and flat--bottomed profile of the lines.
The Si\,{\sc ii} $\lambda\lambda$5958,5979\AA\ feature gives an expansion velocity 
of $\sim 14,300$ km s$^{-1}$,  and if we assume the line at 4395\AA\ to be 
Si\,{\sc iii} $\lambda 4565$\,\AA, we obtain a lower velocity: 11,200 km s$^{-1}$
(to be compared with the $\sim$17,000 km s$^{-1}$ of the Si\,{\sc ii} $\lambda 
6355$\,\AA).
If the Si\,{\sc iii} identification is correct, it could be  proof of  high 
ionization temperature  in the ejecta.
The Mg\,{\sc ii} $\lambda 4481$\,\AA\  minimum gives an expansion velocity of 
17,900 km s$^{-1}$, similar to the other HVFs.
The S\,{\sc ii}  $\lambda\lambda$5454,5640\AA\ 
(the typical ``W'' feature) gives a velocity of about 12,000 km s$^{-1}$,
significantly lower than what is measured for the Si\,{\sc ii} lines.
The evolution  of this feature with time is shown 
 in Fig. \ref{SiIIspecplotcal}; it can be followed until $\sim$4 days
after maximum and  has  disappeared in the next  available spectrum taken 
about  2 weeks after maximum.
The lower velocities measured for Si\,{\sc iii} $\lambda 4565$\,\AA\ and S\,{\sc ii}
 $\lambda\lambda$5454,5640\AA\ suggest that these ions are mainly 
present in a  lower--velocity region with respect to Si\,{\sc ii}, Ca\,{\sc ii}, O\,{\sc i}, and
Mg\,{\sc ii}.
The S\,{\sc ii} $\lambda$5454,5640\AA\ velocity evolution is shown in Fig. 
\ref{SIIVelocity}. 
Interestingly enough in this plot, SN 2004dt
is not one of the extreme objects as in the previous plots 
(Fig. \ref{SiIIVelocity}, \ref{CaHKVelocity}), and it is quite similar to 
SN 2002er, and SN 1994D (see also Fig. \ref{SN2004dt1994D}).

The Ca\,{\sc ii} IR triplet gives a velocity of about 16,500 km s$^{-1}$,
but there is a hint of a bluer component whose minimum at  $\sim 8000$\,\AA\,
corresponds to an expansion velocity of 21,000 km s$^{-1}$.
These values can be compared to the velocity measured for the 
two components and at similar phases for
SN 2002dj (17,700-27,600 km s$^{-1}$ at -11 days), 
SN 2001el (17,100-23,800  km s$^{-1}$ at -9 days), 
SN 2003du (15,500-22,500 km s$^{-1}$ at -11 days), 
SN 2003kf (12,600-23,500 km s$^{-1}$ at -9 days), 
SN 2002er (15,600-23,100 km s$^{-1}$ at -7 days),
SN 2002bo (14,900-22,100 km s$^{-1}$ at -8 days), and
SN 2003cg (12,700-22,000 km s$^{-1}$ at -8.5 days) (data from 
Mazzali et al. \cite{Mazz05a}). Like SN 2004dt, all  these objects show HVFs
in the Ca\,{\sc ii} IR triplet. HVFs are  best detected  in very early spectra 
(earlier than $\sim 1$ week before $B$ maximum) 
in the Ca\,{\sc ii} IR triplet but also in the Si\,{\sc ii} $\lambda$6355\AA\ 
line. They are supposed to arise from abundance or density enhancements (see Sect.  \ref{density}).
\\
\\

\begin{figure}[hbtp]
\includegraphics[angle=0,width=\columnwidth]{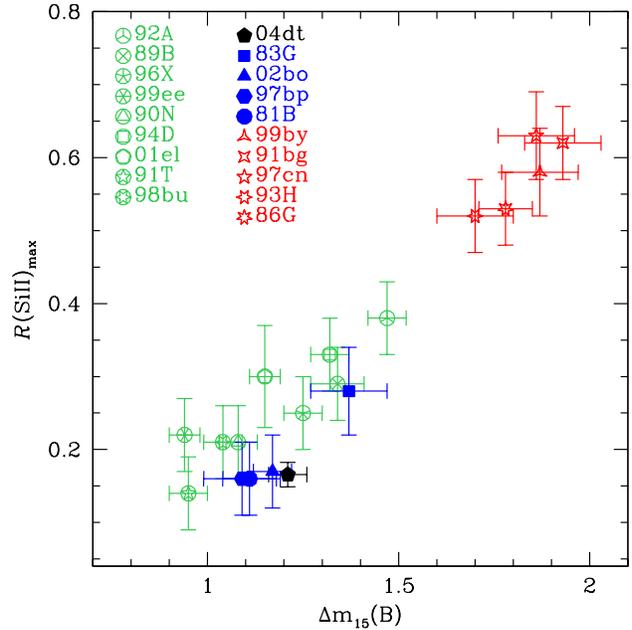}
\caption{ $\mathcal{R}$(Si\,{\sc ii}) vs. $\Delta m_{15}B$  
for SN 2004dt and a sample of SNe shown for comparison.
Filled symbols refer to HVG SNe, starred symbols to FAINT SNe, and all the 
others to LVG SNe, as defined in  Benetti et al. (\cite{Bene05}).
}
\label{itpRSiIIdm15pointnames2004dt}
\end{figure}

\begin{figure}[hbtp]
\includegraphics[angle=0,width=\columnwidth]{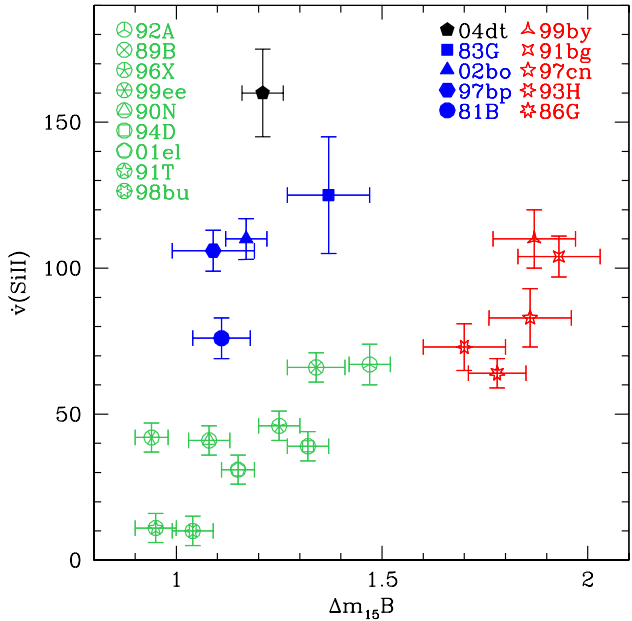}
\caption{ $\dot v$ vs. $\Delta m_{15}B$  
for SN 2004dt and a sample of SNe shown for comparison.
Filled symbols refer to HVG SNe, starred symbols to FAINT SNe, and all the 
others to LVG SNe,
as defined in  Benetti et al. (\cite{Bene05}).
}
\label{dm15vpunto}
\end{figure}

As shown by Nugent et al. (\cite{nugent1995}), there is a correlation between
the ratio  $\mathcal{R}$(Si\,{\sc ii}) of the depth of the Si\,{\sc ii} $\lambda$5972\AA\ and Si\,{\sc ii} 
$\lambda$6355\AA\ absorption troughs near maximum with the decline speed of the B light
curve (i.e. the luminosity), in the sense that higher values of 
$\mathcal{R}$(Si\,{\sc ii}) correspond to subluminous, fast--declining events
(but the correlation fails for $\Delta m_{15}(B)\lesssim 1.2$,  
Benetti et al. \cite{benetti2004}). We measured $\mathcal{R}$(Si\,{\sc ii})$\sim 0.16$ (day +1). 
This value, together with the estimate of $\Delta m_{15}(B)= 1.21$
falls    within   the HVG supernova area 
 in the $\mathcal{R}$(Si\,{\sc ii})$-\Delta m_{15}(B)$ plot
(Fig. \ref{itpRSiIIdm15pointnames2004dt}, see also Benetti et al. \cite{Bene05}).
This result is also confirmed  by the   $\dot v - \Delta m_{15}(B)$ plot
(Fig. \ref{dm15vpunto}, see also Benetti et al. \cite{Bene05}) in which the HVG, 
low--velocity gradient (LVG) and faint SNe are
separated better than in  Fig. \ref{itpRSiIIdm15pointnames2004dt}.

\begin{figure}[hbtp]
\includegraphics[angle=0,width=\columnwidth]{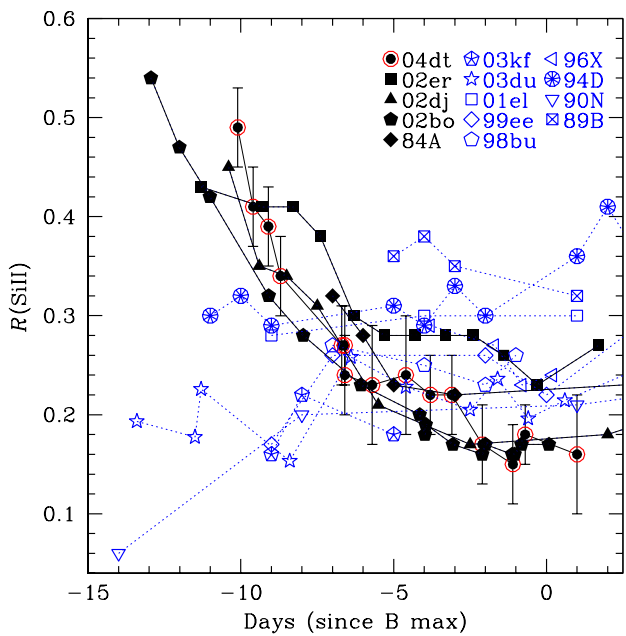}
\caption{Premaximum temporal evolution of the  $\mathcal{R}$(Si\,{\sc ii}) parameter
for SN 2004dt and a sample of SNe shown for comparison.
Filled symbols refer to HVG SNe, open symbols to LVG SNe,
as defined in  Benetti et al. (\cite{Bene05}).
}
\label{RSiII}
\end{figure}

\begin{figure}[hbtp]
\includegraphics[angle=0,width=\columnwidth]{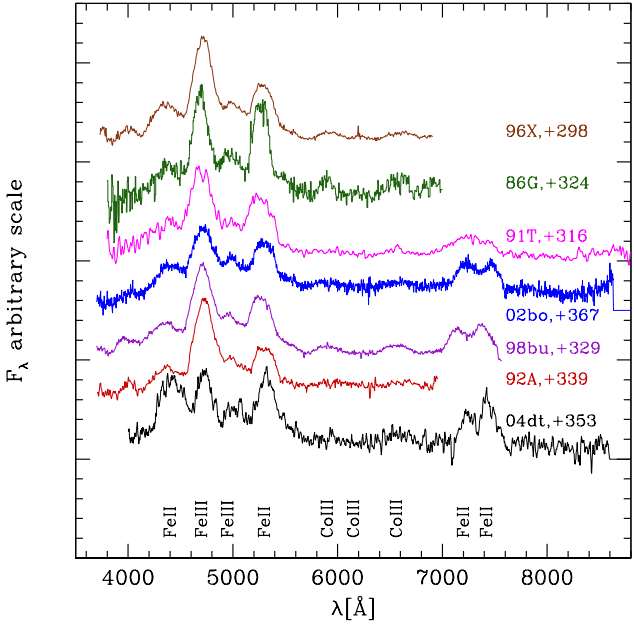}
\caption{Comparison of late spectra ($\sim 1$ year past maximum) 
of different SNe,
including the peculiar SN 1991T and  SN 1986G.
At this phase, SN 1991T and SN 1998 bu have  not yet been  affected by the light
echo contamination (Schmidt et al. \cite{schmidt1994}; 
Cappellaro et al. \cite{cappellaro2001}).
Spectra have been normalized and shifted   for clarity. 
The phase with respect to the  maximum of the B light curve 
is shown close to the name. SN 2004dt spectrum has been slightly smoothed.}
\label{comparison}
\end{figure}

This plot shows that SN 2004dt has the highest  $\dot v$ in our sample
($\dot v=160\pm15$)\footnote{the most similar obiect to SN 2004dt is SN 1983G
($\dot v=125\pm20$,   Benetti et al. \cite{Bene05}). More recent measurements
show that HVG SNe  may have $\dot v$ well above 100, and in particular 
SN 1983G is characterized by   $\dot v\sim 148$ (Benetti, private 
communication), slighlty higher than the value reported in the plot, 
and closer to SN 2004dt value.}. Also, the  evolution of
 $\mathcal{R}$(Si\,{\sc ii}) (Fig. \ref{RSiII})  is very similar to the HVG
 SNe as shown in  Benetti et al. (\cite{Bene05}).
In conclusion, SN 2004dt seems to belong to the ``high--velocity'' SNe Ia,
such as SN 2002bo (Benetti et al. \cite{benetti2004}), SN 2002dj (Pignata et al. \cite{pignata2005})  or SN 2002er 
(Kotak et al. \cite{kotak2005}; Pignata et al. \cite{pignata2004}).
Nevertheless, SN 2004dt also shows some peculiarities with respect to the HVG SNe
(Fig. \ref{spec04dt96X}, Fig. \ref{SiIIspecplotcal} left panel), 
in particular the  Si\,{\sc ii}  ($\lambda 6355$) feature
remains broader and probably visible for
a longer time than for most of HVG SNe ($>28$d).
However, even if they are kinematically different from normal SNe Ia, 
their photometric properties, such as the peak luminosity and the light--curve 
decline ($\Delta m_{15}$), do not show any peculiarity, although the color
evolution can be affected, as mentioned in Sect. \ref{colourevolution}
(see also Benetti et al. \cite{benetti2004}). 
\\
\\
Even if the present work is focused on the early spectral evolution, 
it is worth short  comment here  on the very last 
spectrum in our sample, taken almost  one year after maximum (day  +353).
At this phase,  when the ejecta  have become  thin and transparent 
to radiation, spectra are usually  dominated by the strong 
emission lines of Fe\,{\sc [ii]}  and Fe \,{\sc [iii]} originating from the
radioactive decay of $^{56}$Ni into stable $^{56}$Fe through $^{56}$Co  
(Kuchner \cite{kuchner1994}).
\\
 Figure \ref{comparison}   shows the comparison of a  late spectrum 
of SN 2004dt  with
spectra of other SNe at a similar phase
(SN 1996X, Salvo et al. \cite{salvo2001};
 SN 1986G,  Cristiani et al. \cite{cristiani1992};
 SN 1991T, G\'{o}mez \& L\'{o}pez \cite{gomez1998};
 SN 1998bu, Cappellaro et al. \cite{cappellaro2001};
 SN 1992A,  ESO+Asiago SN archive;
 SN 2002bo, ESO+Asiago SN archive, Benetti et al. \cite{benetti2004}). 
In spite of the paucity of 
Fe--group  elements observed at early epochs, as discussed in Sect. \ref{density}, 
this nebular spectrum apparently shows all the 
expected iron features, suggesting  that iron-group elements were prevalent 
in the inner layers. 
At late phases SN 2004dt,  resembles SN 2002bo 
(Benetti et al. \cite{benetti2004}),
which is in fact another HVG SN, but it also shows some peculiarities
that deserve  further investigation, such as 
the Fe\,{\sc [iii]}/ Fe\,{\sc [ii]} line
ratio, with the  Fe\,{\sc [iii]} $\lambda\sim4700$\AA\ and  Fe\,{\sc [ii]} 
$\lambda\sim5300$\AA\ features with similar peak intensity, and an 
intriguing Fe\,{\sc [ii]} $\lambda\sim4400$\AA\ feature that is  particularly strong.
The late spectral evolution of SN 2004dt
will be  analyzed and discussed in more detail in  a forthcoming paper.

\section{Spectral modeling} \label{density}
The availability  of early SNe Ia spectra,
whose number is increasing thanks to observational efforts,  such as the ones
performed by the  European Supernova Collaboration, shows that the
high--velocity line features are quite common in SNe~Ia
(Mazzali et al. \cite{Mazz05a};
Benetti et al. \cite{Bene05};
Garavini et al. \cite{garavini2007}),
but most objects only show high--velocity
components  in some lines, preferentially in Si\,{\sc ii} and Ca\,{\sc ii}
lines. SN~2004dt seems to be an extreme case. All lines of the early
time spectra, except for sulfur, have prominent high--velocity
absorptions. Even O\,{\sc i} is strongly affected, which is very unusual.
In order to find physical explanations of these aspects from a more
theoretical point of view, spectral synthesis calculations were
performed. We used a Monte Carlo code to calculate synthetic
spectra. A detailed description of the code can be found in
Abbott \& Lucy (\cite{Ab85}), Mazzali \& Lucy (\cite{ml93}), and Lucy (\cite{l99}).
Another extension of the  Abbott \& Lucy's (\cite{Ab85})
code was developed 
for premaximum spectral analyses of SNe~Ia (Ruiz--Lapuente \cite{ruiz92a}; Ruiz--Lapuente 
et al. \cite{ruiz92b}).
Recently, the Lucy (\cite{l99}) code was enhanced in order to derive the distribution
of abundances in SNe~Ia ejecta allowing detailed supernova tomography
(Stehle \cite{Steh04}; Stehle et al. \cite{Steh05}).

Here the analysis is focused on an early spectrum and on another 
one near
maximum light. The early epochs are best--suited to examining the outer
parts of the ejecta, hence the high--velocity components of
the absorption lines. Since the SN ejecta have not yet expanded very
much, the outer layers are still dense enough to produce significant
absorptions.

Models for both epochs are initially based on the W7 density profile
(Nomoto et al. \cite{nomoto1984}), which is taken as the standard 
departure model.
The bolometric luminosity, the radius of the inner
boundary (``photosphere''), and the abundances are determined
iteratively  to achieve the best match to the observed spectrum.

\subsection{Standard density profile}\label{standarddensityprofile}
The spectrum at day -9.6 is chosen for modeling  the early epoch.
As shown in  Fig. \ref{SN2004dtspec},
it is the earliest spectrum that
covers a large  wavelength interval, from the UV to the near IR. Input 
distance ($\mu = 34.61$) was taken from the Lyon-Meudon Extragalactic 
Database\footnote{http://cismbdm.univ-lyon1.fr/$\sim$leda/}, other
input parameters, like the  time since explosion ($t_{exp} = 9.1$\,d)
and reddening ($E(B-V) = 0.00$\footnote{due to the discrepancies in the
observed reddening discussed in Sect. \ref{colourevolution} and the absence
of clearly detectable Na\,{\sc i}, we assumed the reddening to be 
negligible.}),
 were derived from the observations.
\begin{figure}
\centering
\includegraphics[angle=270,width=\columnwidth]{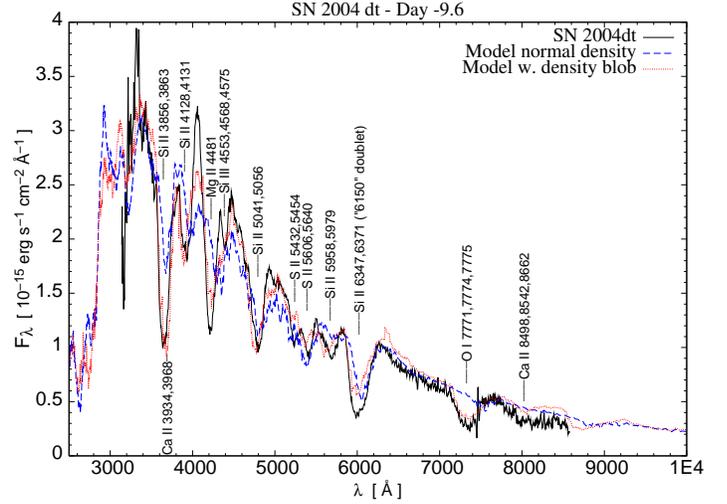}
\caption{Observed spectrum and corresponding model of SN~2004dt at
day $-9.6$ using $E(B-V)=0.0$, $\mu = 34.61$, $t_{exp} = 9.1$\,d.}
\label{min9_8}
\end{figure}
The  dashed line in Fig.~\ref{min9_8} shows the best--fitting  model. The
derived luminosity is log$_{10}L = 42.87$\,(erg\,s$^{-1}$), and the photospheric
radius is located at 14,500\,km\,s$^{-1}$. To account for the high--velocity 
components, three shells in the outer layers --- at 18,000,
20,000, and 23,000\,km\,s$^{-1}$ --- were introduced. These
velocities correspond to the main regions where different lines were detected.

Two discrepancies between the model and the observations are evident: 
first, except for the S\,{\sc ii} W--feature, all synthetic lines are redder than 
the observed ones, or at least lack absorptions in their blue wings. 
Second, there is almost no Si\,{\sc ii} absorption at 5970\,\AA, and the Si\,{\sc iii} line 
at 4560\,\AA\ is too strong, which suggests that the model temperature is too 
high. The Si\,{\sc iii} lines are formed near the photosphere, where the temperature is 
very high, while Si\,{\sc ii} lines are formed farther out, where the temperature is 
lower. 
Using a standard density profile
leaves too little mass in regions above $\sim 15,000$\,km\,s$^{-1}$ to
produce significant absorptions. Oxygen is singly ionized at low
velocities, but the strong O\,{\sc i} line at 7772\,\AA\ is absent in the
model, due to the low density at velocities $\sim 17,000$\,km\,s$^{-1}$,
where its main observed absorption occurs. Only the S\,{\sc ii} W-feature
($\sim 5630$\,\AA) can be reproduced fairly well. These lines are
comparably weak and are formed near the photosphere. Therefore, they
do not show significant high--velocity absorption.

\begin{figure}
\centering
\includegraphics[angle=270,width=\columnwidth]{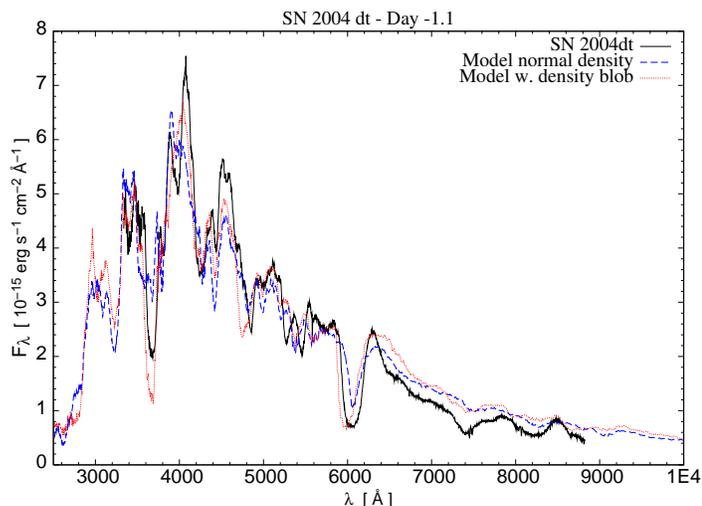}
\caption{Observed spectrum and corresponding model of SN~2004dt at
day $-1.1$ using $E(B-V)=0.0$, $\mu = 34.61$, $t_{exp} = 17.7$\,d.}
\label{min1_3}
\end{figure}

The maximum light spectrum was assumed to have an epoch from explosion of $t_{exp}
= 17.7$\,d. The photospheric velocity had decreased to 9200\,km\,s$^{-1}$, so
the model luminosity was determined to be log$_{10}L = 43.04$\,(erg\,s$^{-1}$).
Figure~\ref{min1_3} 
shows that the original W7 model (dashed line)  behaves as at
the early time. The overall flux shape is  reproduced well, but most
lines are too weak  compared with the observed spectrum. Again, the
temperature near the model photosphere is high enough that Si, Mg,
and Fe are doubly ionized. Therefore, the transitions of singly
ionized species are too weak in the synthetic spectrum, 
seen best in  Si\,{\sc ii} $\lambda$3860\,\AA, 4130\,\AA, and 6355\,\AA. Consequently, the
Si\,{\sc iii} $\lambda$4560\,\AA\ line is too strong  compared with the observed one.

The two spectral synthesis models show that the
deep, blue--line absorption cannot be reproduced by simply
increasing the abundances, especially because HVFs
affect not just a few lines but almost all lines --- including O\,{\sc i} $\lambda$7772\AA. 
Since more absorption at higher
velocities is necessary and increasing the abundances in the outer
regions is not sufficient, the other possibility is to increase the
density. Plausible reasons for an enhanced density are be discussed
later.

\subsection{Enhanced density profile}
To improve the models, the density at high velocities was increased 
in steps until the line absorptions could be reproduced both in
depth and velocity. Furthermore, attention was paid to keeping the
density profile smooth and similar to the original shape of the W7 
profile.
We also tried to keep the abundances at reasonable values,
i.e. similar to the values that apply to other ``normal''  SNe~Ia. The
best match was achieved by adding 0.26\,M$_{\odot}$ above 16,000\,km\,s$^{-1}$ in a
spherically symmetric shell.

The dotted line in Fig.~\ref{min9_8} shows the model, using the same input 
parameters and abundances as before, but with the higher density. The 
improvements compared to the model without a density enhancement are striking.
The combined feature of Ca\,{\sc ii},H\&K, and Si\,{\sc ii} $\lambda$3856\,\AA, as well as
the Si\,{\sc ii} $\lambda$4130\,\AA\ line, are reproduced very well, both in
absorption depth and velocity. The Mg\,{\sc ii} $\lambda$4481\,\AA\ line is somewhat
too shallow, but now it is at the correct wavelength. Si\,{\sc iii} at 
$\lambda$4560\,\AA\ is strongly reduced compared with the low--density model,
while the Si\,{\sc ii} $\lambda$5045\,\AA\ line is deeper and fits the observations
very well. The S\,{\sc ii} absorption between 5090\,\AA\ and 5660\,\AA\ is 
slightly weaker, but it still is at the right place, 
and a significant absorption due to Si\,{\sc ii} $\lambda$5960\,\AA\ is visible.
Lower temperatures in the outer layers, in comparison with those near the 
photosphere, allow this line to be formed. The main Si feature, Si\,{\sc ii} at 
$\lambda$6350\,\AA, is still not deep enough, but the model shows the 
high--velocity component that is not present in the low--density model. The
effect of an increased density can be seen nicely for O\,{\sc i} $\lambda$7772\,\AA. 
This line was completely absent in the model without a density enhancement, 
but now the model matches the observed line profile almost perfectly.

The model near maximum light shows a similar change when the density is
increased in the same way 
as for the model at the early epoch. All lines are shifted
bluewards. The Si\,{\sc ii} lines at 3856\,\AA\ and 6350\,\AA\ are
comparable to the observations. Fe\,{\sc iii} at 4420\,\AA\ comes in and
dominates the feature observed at $\sim 4250$\,\AA. The Si\,{\sc iii}
complex at 4560\,\AA\ is still a little bit too strong but represents
the observed line much better. A fairly strong absorption near
4750\,\AA\ is caused by Si\,{\sc ii} 5041, 5056\,\AA. In the model it
appears to be formed at  velocities that are too high. The same applies to the
main Si\,{\sc ii} feature at 6350\,\AA, which is blue--shifted with respect
to the observations. Even with an increased density, there is not
enough absorbing mass to reach the depth of the O\,{\sc i} $\lambda$7772\,\AA\ line.
Also the Ca\,{\sc ii} IR-triplet is underestimated in the model. A
higher Ca~abundance, however, would also increase the absorption 
depth of Ca\,{\sc ii},H\&K, which is already too strong.

From the analysis using synthetic spectra, it can be concluded that
SN~2004dt contains an unusually high Si abundance, since almost
every prominent feature in the spectra is dominated by transitions
of this element. This applies not only to the early epochs but also
to the spectrum near maximum light, when usually Si high--velocity
components tend to disappear. Even the deep feature near 3700\,\AA,
which is normally Ca\,{\sc ii},H\&K in most SNe~Ia, is dominated by Si\,{\sc ii}
transitions. On the other hand, only very little Fe-group elements
are found in SN~2004dt. The outer layers ($14,500$\,km\,s$^{-1}$$\le v \le
20,000$\,km\,s$^{-1}$) contain 0.5\% by mass of Fe-group elements, and even in
deeper layers between 9200\,km\,s$^{-1}$ and 14,500\,km\,s$^{-1}$ only 12\% can be
found, where  $^{56}$Ni and $^{56}$Fe begin to dominate in other SNe~Ia.
\\

The analysis of the observations has already shown the significant 
high--velocity absorptions in almost every line except for the S\,{\sc ii} and
Si\,{\sc iii} lines. Modeling those spectra has confirmed that, and proved
that models with classical density profiles do not contain enough material 
in their outer regions to reproduce the observations. 
Objects like SN~2004dt clearly show in the spectral
features the effects of 3-D structures in the expanding ejecta, and so, they
are a very  strong hint of aspherical explosion scenarios.

\section{Discussion and conclusions}

SN~2004dt clearly differs from most `normal' SNe~Ia with respect to line 
velocities. The observed high--velocity  absorption can be reproduced in 
synthetic spectra only when the density is increased significantly above 
$\approx 15,000$\,km\,s$^{-1}$ with respect to models that fit
the spectra of less extreme HVG SNe (Stehle et al. 2005).
Global enhancement may be a property of the explosion or the result of
interaction with circumstellar material (CSM) (Gerardy et al. \cite{gerardy2004}; 
Mazzali et al. \cite{Mazza05b}). In 
both cases, a spherically symmetric situation is unlikely: a density enhancement
would require a mass $\approx 0.26$\,M$_{\odot}$, while  the added mass in the
case of interaction would be even larger since the CSM would 
probably be dominated by elements such as H and/or He. 
A possibility is that the outer ejecta are shaped by 
rising finger-like structures produced in the burning and that these may
interact with CSM. The ubiquitous presence of HVFs in the earliest spectra of
SNe Ia (Mazzali et al. \cite{Mazz05a}) was reproduced by 3-D 
spectrum synthesis models assuming a limited number of blobs (6 to 12) of 
large angular size at $v \sim 18000-25000$ \,km\,s$^{-1}$ (Tanaka et al. 
\cite{tanaka06}).
If one such blob happened to cover the early-time photosphere of SN2004dt 
almost completely, the observed line profiles may be recovered, while the total
mass of high-velocity material would not exceed $\sim 0.1$\,M$_{\odot}$.

The idea of finger-like or blob structures can probably be extended to 
all HVG SNe.
SN 2004dt is a member of a group of SNe~Ia with unusually 
high--velocity absorption lines, indicating ejecta moving 
4000--5000 km s$^{-1}$ faster than the velocities typically measured
in most SNe~Ia at the same phases (Leonard et al. \cite{leonard2005}). 
Other SNe in this class include SN 1981B, SN 1983G, SN 1984A, SN 1989B, 
SN 1991M, SN 1995al, SN 1997bp, SN 2002bo, SN 2002bf, SN 2002dj, SN 2002er 
(see also Benetti et al. \cite{Bene05}), 2005cf (Pastorello et al. 
\cite{pastorello2007}; Garavini et al. \cite{garavini2007}). In the case of 
SN 2002bf, the spectropolarimetric data of Leonard et al. (\cite{leonard2005}) 
show similarities with SN 2004dt.

From the analysis of the spectropolarimetry, Leonard et al. (\cite{leonard2005})
favor clumpiness of the outer layers.
Clumpiness of the ejecta is also supported by Wang, Baade \& Patat (\cite{wang2007}),
 who find that the greater the line polarization (i.e. clumpiness), 
the faster the decline and the less luminous the SN.
Due to the clumpy structure of the ejecta, projection effects
may alter the observed polarization, suggesting that the peak 
luminosity/decline rate relationship in the light curves of SNe~Ia could be 
explained by the varying clumpiness of the ejecta, combined with the viewing angle.
Remarkably, SN 2004dt deviates significantly from the correlation between the
degree of polarization across the Si\,{\sc ii} $\lambda$6355\AA\ line 
($P_{SiII}$) and the decline
rate $\Delta m_{15}(B)$ found by Wang, Baade \& Patat (\cite{wang2007}), confirming its
peculiarity. The same authors suggest that projection effects may account
for this deviation.
Clumpiness has also
been observed in the remnant of Tycho's SN (Badenes et al. \cite{badenes2005}).
Although based on just 1-D modeling of
the ejecta, our own results point in the same direction. 

It is still unclear what type of explosion is most suitable for 
producing intrusions of intermediate--mass elements moving at high 
velocities into (or above) outer layers of unburned O and C. 
Subsonic burning, the defining characteristic of deflagration models 
of the explosions, can reproduce such mixing, whereas detonation models are 
thought to burn the outer layers without producing clumps. A 
possible exception to the latter could be the model of Plewa et al.
(\cite{plewa2004}), where the intrusion of high--velocity, intermediate--mass 
elements would come from the same rising plume that triggers a detonation 
in the outer layers (Kasen \& Plewa \cite{kasen2005}).  However, the presence 
of a single blob would be at odds with the presence of HVFs in all SNe Ia.

The combination of the spectroscopic and photometric observations
of SN 2004dt with 1D modeling of the ejecta and of the corresponding
spectra  presented here strengthens the conclusions from independent
spectropolarimetric analyses of this and similar SNe~Ia. It equally
stresses the need for realistic 3-D models of the explosion to deal
with current and future data,  and also the importance of early
observations, both spectroscopic and spectropolarimetric.

\begin{acknowledgements}
This work has been supported by the European Community's Human 
Potential Program under contract HPRN-CT-2002-00303, ``The Physics
of Type Ia Supernovae''.
\\
This work is based on observations collected at 
the 2.3m telescope at the Siding Spring Observatory (SSO, Australia),
at the 2.2m telescope at the  Centro Astron\'omico Hispano Alem\'an 
(CAHA, Spain), 
at the 3.6m Telescopio Nazionale Galileo (TNG, La Palma, Canary Islands), 
at the 2.5m Nordic Optical Telescope (NOT, La Palma, Canary Islands), and 
at the 1.8m telescope at the Asiago Astrophysical Observatory (Italy).
The SSO is operated by the Australian National University.
The CAHA at Calar Alto is operated jointly by the Max-Plank Institut
f\"ur Astronomie and the Instituto de Astrof\'{\i}sica  de Andaluc\'{\i}a 
(IAA-CSIC).
The TNG is operated on the island of La Palma 
by the Centro Galileo Galilei of the Istituto Nazionale di Astrofisica
(INAF), in the Spanish Observatorio del
Roque de Los Muchachos  of the Instituto de 
Astrof\'{\i}sica de Canarias.
The NOT is operated  on the island of La Palma 
jointly by  Denmark, Finland, Iceland, Norway, and Sweden,
in the Spanish Observatorio del
Roque de Los Muchachos  of the Instituto de 
Astrof\'{\i}sica de Canarias.
The Astrophysical Observatory of Asiago is operated by the Padua Observatory 
(member of the National Institute for Astrophysics INAF) and by the Astronomy 
Department of the University of Padua.
\\
We would like to thank J. N\"ar\"anen, K. Muinonen, X. Wang, M. Weidinger, R. Karjalainen, and J. Telting for observing SN 2004dt at NOT.
We are  grateful to all the other  observers and to the staff at the various
observatories for efficiently scheduling and carrying out our requests and to
the observers who had to give up part of their observing time so that
SN 2004dt could be observed.
\\ 
This research made use of the NASA/IPAC Extragalactic Database (NED), 
which is operated by the Jet Propulsion Laboratory, California Institute of 
Technology, under contract with the National Aeronautics and Space 
Administration. We have also made use of the Lyon-Meudon Extragalactic Database
(LEDA), supplied by the LEDA team at the Centre de Recherche Astronomique 
de Lyon, Observatoire de Lyon.
\end{acknowledgements}


\clearpage



\begin{thebibliography}{}



\bibitem[1985]{Ab85} 
Abbott,~D.C., \& Lucy,~L.B. 1985, ApJ, 288, 679
\bibitem[2004]{altavilla04}
Altavilla, G., et al.
2004, MNRAS, 349, 1344
\bibitem[2005]{badenes2005}
Badenes, C., Borkowski, K.J., Hughes, J.P., Hwang, U., \& Bravo, E.
2006, ApJ, 645, 1373


\bibitem[2004]{benetti2004}
Benetti, S., et al. 2004, MNRAS, 348, 261
\bibitem[2005]{Bene05}
Benetti,~S., et al.
2005, ApJ, 623, 1011



\bibitem[2003]{branch2003}
Branch, D.,  Garnavich, P.,  Matheson, T.,  Baron, E.,  Thomas, R.C.,  Hatano, K.,  Challis, P.,   Jha, S.,  Kirshner, R.P. 2003, AJ, 126, 1489 

\bibitem[1982]{burstein1982} 
Burstein, D., \&  Heiles, C. 1982, AJ, 87, 1165

\bibitem[2001]{cappellaro2001}
Cappellaro, E., Patat, F., Mazzali, P.A., Benetti, S., Danziger, J.I., Pastorello, A., Rizzi, L., Salvo, M., Turatto, M. 2001, ApJ, 549, 215

\bibitem[1992]{cristiani1992} 
Cristiani, S., et al.
1992, A\&A, 259, 63



\bibitem[2001]{filippenko2001}
Filippenko, A.V., Li, W.D., Treffers, R.R.,  Modjaz, M. 2001, ASPC, 246, 121



\bibitem[2003]{gamezo2003}
Gamezo, V.N., Khokhlov, A.M., Oran, E.S., Chtchelkanova, A.Y., \& 
Rosenberg, R.O. 2003, Science, 299, 77

\bibitem[2007]{garavini2007}
Garavini, G., et al. 2007, A\&A, 471, 527  


\bibitem[2004]{gerardy2004}
Gerardy, C.L.,  H\"{o}flich, P., Fesen, R.A., Marion, G.H.,  Nomoto, K., Quimby, R., Schaefer, B.E., 
Wang, L., Wheeler, J.C.,  2004, ApJ, 607, 391



\bibitem[1998]{gomez1998}
G\'{o}mez, G. \& L\'{o}pez, R. 1998, AJ, 115, 1096



\bibitem[2000]{hillebrandt2000}
Hillebrandt, W., \& Niemeyer, J.C. 2000, ARA\&A, 38, 191



\bibitem[1986]{horne1986}
Horne K. 1986, PASP, 98, 609


\bibitem[2005]{kasen2005}
Kasen, D., \& Plewa, T. 2005, ApJ, 622, 41

\bibitem[1991]{khoklov1991}
Khoklov, A. 1991, A\&A, 245, 114


\bibitem[1993]{kirshner1993}
Kirshner, R.P.,  et al.
1993, ApJ, 415, 589


\bibitem[2005]{kotak2005}
Kotak, R., et al. 
2005, A\&A, 436, 1021




\bibitem[1994]{kuchner1994}
Kuchner, M.J., Kirshner, R.P., Pinto, P.A., \& Leibundgut, B. 1994 
ApJ, 426, L89






\bibitem[2000]{lentz2000} 
Lentz, E.J., et al.
2000, ApJ, 530, 966

\bibitem[2005]{leonard2005}
Leonard, D.C., Li, W., Filippenko, A.V., Foley, R.J., \& Chornock, R. 2005,
ApJ, 632, 450

\bibitem[1995]{Lira1995} 
Lira,  P. 1995, Masters thesis, Univ. Chile


\bibitem[1999]{livne1999}
Livne, E. 1999, ApJ, 527, L97

\bibitem[1999]{l99} 
Lucy, ~L.B., 1999,A\&A, 345, 211

\bibitem[2003]{lundqvist2003}
Lundqvist, P.,  Sollerman, J., Leibundgut, B.,  Baron, E.,  Fransson, C.,   Nomoto, K.,  in  
From Twilight to Highlight: The Physics of Supernovae: Proceedings of the ESO/MPA/MPE Workshop Held at Garching, Germany, 29-31 July 2002, ESO ASTROPHYSICS SYMPOSIA. ISBN 3-540-00483-1.  W. Hillebrandt and B. Leibundgut eds. Springer-Verlag, 2003, p. 309


\bibitem[1989]{marsh1989}
Marsh T. 1989, PASP, 101, 1032

\bibitem[2000]{maxted2000}
Maxted, P.F.L., Marsh,  T.R., \& North  R.C. 2000, MNRAS, 317, L41 


\bibitem[2001]{mazzali2001}
 Mazzali, ~P.A. 2001, MNRAS, 321, 341

\bibitem[1993]{ml93}
Mazzali,~P.A., \& Lucy~L.B. 1993, A\&A, 279, 447



\bibitem[2005a]{Mazz05a} Mazzali,~P.A., et al.
  2005a, ApJL, 623, 37

\bibitem[2005b]{Mazza05b}  Mazzali,~P.A., et al.
2005b, MNRAS, 357, 200

\bibitem[2007]{mazzali2007} Mazzali,~P.A., R\"{o}pke, F.K., Benetti, S., Hillebrandt, W.  2007, Science, 315, 825

\bibitem[2004]{moore2004}
Moore, M., \& Li W. 2004, IAUC 8386
\bibitem[2004]{napiwotzki2004}
Napiwotzki, R., et al. 2004, RMxA\&A, 20, 113



\bibitem[1984]{nomoto1984}
Nomoto,~K., Thielemann, F.-K., \& Yokoi, K. 1984, ApJ, 286, 644

\bibitem[1995]{nugent1995}
Nugent,  P., et al.
1995, ApJ, 455, 147


\bibitem[1997]{nugent1997}
Nugent, P., Baron, E., Branch, D., Fisher, A., \&  Hauschildt, P.H. 
1997, ApJ, 485, 812

\bibitem[2007]{pastorello2007}
Pastorello, A., Taubenberger, S., Elias-Rosa, N., et al. 2007, MNRAS, 376, 1301  



\bibitem[1996]{patat1996}
Patat, F., et al. 1996 MNRAS 278, 111
\bibitem[2004]{patat2004}
Patat, F., Pignata, G., \& Benetti, S. 2004, IAUC 8387
\bibitem[1999]{perlmutter1999}
Perlmutter, S., et al. 1999, ApJ, 517, 565 



\bibitem[2004]{pignata2004}
Pignata, G., et al.
2004, MNRAS, 355, 178


\bibitem[2005]{pignata2005}
Pignata, G., et al.
 in Turatto M,  Benetti S.,  Zampieri L.,  Shea W., eds, ASP Conf. Ser. Vol. 342, 1604-2004: Supernovae as Cosmological Lighthouses. Astron. Soc. Pac., San Francisco, p. 266, 2005.


\bibitem[2004]{plewa2004}
Plewa, T., Calder, A.C., \& Lamb, D.Q. 2004, ApJ, 612, L37


\bibitem[1969]{pskovskii1969} 
Pskovskii, Y.P.  1969, SvA, 12, 750

\bibitem[2002]{reinecke2002}
Reinecke, M., Hillebrandt, W., \& Niemeyer, J.C. 2002, A\&A, 391, 1167
\bibitem[1998]{riess1998}
Riess, A., et al. 1998, AJ, 116, 1009  


\bibitem[2006]{ropke2006}
R\"{o}pke, F.K., Woosley, S.E. \& Hillebrandt, W. 2007, ApJ, 660, 1344 

\bibitem[1992a]{ruiz92a}
Ruiz--Lapuente, P. 1992a, PhD Thesis, Univ. of Barcelona
\bibitem[1992b]{ruiz92b}
Ruiz--Lapuente, P., et al. 1992b, ApJ, 387, L33
\bibitem[2004]{ruiz2004}
Ruiz--Lapuente, P., et al. 2004, Nature, 431, 1069

\bibitem[2001]{salvo2001}	
Salvo, M.E., Cappellaro, E., Mazzali,  P.A., Benetti, S., Danziger, I.J., Patat, F.,  \& Turatto, M. 2001, MNRAS, 321, 254

\bibitem[2004]{salvo2004}
Salvo, M.E., Schmidt B., \& Wood P. 2004, IAUC 8387


\bibitem[1998]{schlegel}
Schlegel, D.J.,  Finkbeiner D.P., \& Davis M. 1998, ApJ, 500, 525


\bibitem[1994]{schmidt1994}
Schmidt, B.P., Kirshner, R.P., Leibundgut, B., Wells, L.A., Porter, A.C., Ruiz-Lapuente, P., Challis, P., Filippenko, A.V. 1994, ApJ, 434, 19


\bibitem[2006]{stanishev2006} 
Stanishev, V., et al. 2006, submitted to A\&A
\bibitem[2004]{Steh04}
Stehle,~M. 2004, PhD Thesis, Ludwig--Maximillians-Univ.
\bibitem[2005]{Steh05}
Stehle,~M., Mazzali,~P.A., Benetti,~S., \& Hillebrandt~W. 2005, MNRAS,
360, 1231



\bibitem[2006]{tanaka06}
Tanaka, M.; Mazzali, P.A.; Maeda, K.; Nomoto, Ken'ichi 
2006, ApJ, 645, 470

\bibitem[1986]{thielemann1986}  
Thielemann,  F.-K., Nomoto, K., \& Yokoi, K. 1986, A\&A, 158, 17
\bibitem[1998]{theureau1998}
Theureau, G., Bottinelli, L., Coudreau-Durand, N., Gouguenheim, L., Hallet, N.,
 Loulergue, M., Paturel, G., \& Teerikorpi, P. 1998, A\&AS, 130, 333


\bibitem[2007]{thomas2007}
Thomas, R.C., et al. 2007,ApJ, 654, 53



\bibitem[2006]{Wang06} Wang,~L.,
Baade,~D., H\"oflich,~P., Wheeler,~C.~J., Kawabata,~K., Khokhlov,~A.,
Nomoto,~K., \& Patat~F. 2006, ApJ, 653, 490



\bibitem[2007]{wang2007} Wang,~L.,  Baade, D. \& Patat, F.
2007, Science,  315, 212
\end{thebibliography}
\end{document}